\documentclass[letterpaper,11pt]{article}

\usepackage{arydshln}
\usepackage{amsmath}
\usepackage{amsthm}
\usepackage{authblk}
\usepackage{setspace}
\usepackage{url}
\usepackage[comma,authoryear]{natbib}
\usepackage{hyperref}
\usepackage{theoremref}
\usepackage[table]{xcolor}
\usepackage{float}
\restylefloat{table}
\usepackage{amssymb}
\usepackage{caption}
\usepackage{subcaption}
\usepackage[shortlabels]{enumitem}
\usepackage{verbatim}
\usepackage[titletoc,title]{appendix}
\setcitestyle{citesep={;}}

\definecolor{armygreen}{rgb}{0.29, 0.33, 0.13}
\definecolor{auburn}{rgb}{0.43, 0.21, 0.1}
\definecolor{burgundy}{rgb}{0.5, 0.0, 0.13}
\definecolor{medium red}{rgb}{.490,.298,.337}
\definecolor{dark red}{rgb}{.235,.141,.161}
\definecolor{dark green}{rgb}{0.0,0.5,0.0}

\hypersetup{
	colorlinks = true,
	linkcolor = {burgundy},
	urlcolor = {burgundy},
	citecolor = {burgundy}, 		
	linkbordercolor = {white},
}

\newtheorem{theorem}{Theorem}
\newtheorem{proposition}{Proposition}
\newtheorem{claim}{Claim}[section]

\theoremstyle{definition}
\newtheorem{example}{Example}

\newcommand*{\claimproofname}{Proof of Claim}
\newenvironment{claimproof}[1][\claimproofname]{\begin{proof}[#1]}{\end{proof}}

\newcommand{\red}{\textcolor{red}}
\newcommand{\blue}{\textcolor{blue}}

\title{Efficient Reallocation of Indivisible Resources: Pair-efficiency versus Pareto-efficiency\thanks{I am grateful to Szilvia P\'{a}pai for helpful discussions. I also appreciate the helpful suggestions provided by three anonymous referees.}}

\author{Pinaki Mandal\thanks{E-mail: \textit{pinaki.mandal@asu.edu}}}

\affil{Department of Economics, Arizona State University, USA}

\date{March 28, 2025}

\begin{document} 	
	\maketitle
	
	\begin{abstract}
		In the object reallocation problem, achieving Pareto-efficiency is desirable, but may be too demanding for implementation purposes. In contrast, pair-efficiency, which is the minimal efficiency requirement, is more suitable. Despite being a significant relaxation, however, pair-efficiency ensures Pareto-efficiency for any strategy-proof and individually rational rule when agents' preferences are unrestricted.
		
		What if agents' preferences have specific restricted structures, such as \textit{single-peakedness} or \textit{single-dippedness}? We often encounter such situations in real-world scenarios. This study aims to investigate whether pair-efficiency is sufficient to ensure Pareto-efficiency in such cases.
		
		Our main contribution in this paper is establishing the equivalence between pair-efficiency and Pareto-efficiency when dealing with single-peaked or single-dipped preference profiles. This equivalence holds without needing to assume any other properties of the rule. We further show that both the single-peaked domain and the single-dipped domain are the ``maximal'' domains where this equivalence holds.
	\end{abstract}
	
	

	\newpage

	\section{Introduction}
	
	This paper considers the \textit{object reallocation problem} \citep{shapley1974cores}: There is a group of agents, each of whom initially owns a distinct indivisible object. Agents have (strict) preferences over all objects. The initial allocation may not be efficient, so the issue of improving efficiency through reallocating the objects (without the use of monetary transfers) arises. Each agent's preference information is private, and a rule specifies how to reallocate objects based on the preference information provided by the agents.
	
	\textit{Pareto-efficiency} is a standard efficiency requirement. It ensures that no group of agents can benefit from trading their assignments without making someone worse off. However, while desirable, Pareto-efficiency may be too demanding for implementation purposes. This is because of the potential persistence of Pareto-inefficient allocations. Even though a Pareto-inefficient allocation allows for efficiency-improving trade, the trade can be hard to identify and coordinate if it involves many agents, and consequently, the allocation may persist after implementation.\footnote{See \citet{feldman1973bilateral} for a similar discussion in a pure exchange economy with divisible resources.}
	
	On the other hand, inefficient allocations are likely to be destabilized if the efficiency-improving trade involves only two agents. Pairwise exchanges are much easier to identify and execute since the involved agents can quickly recognize and act on mutual benefits. This leads to \textit{pair-efficiency}, a natural minimal efficiency requirement that only rules out efficiency-improving trades between pairs of agents, making it more suitable for implementation purposes.
	
	Pair-efficiency is a significant relaxation of Pareto-efficiency. When agents' preferences are unrestricted, there are instances (preference profiles) with $n \geq 7$ agents where there is only one Pareto-efficient allocation, but at least $2^n$ pair-efficient allocations. However, despite being a significant relaxation, pair-efficiency ensures Pareto-efficiency for any \textit{strategy-proof} and \textit{individually rational} rule on the unrestricted domain.\footnote{A rule is \textit{strategy-proof} if no agent can ever benefit from misrepresenting her preferences.}$^{,}$\footnote{A rule is \textit{individually rational} if no agent is ever assigned an object worse than her endowment (the object that she initially owns).} For more details on these results, see \citet{ekici2024pair}.

	\subsection{Our paper}
	
	What if agents' preferences have specific restricted structures? We often encounter such situations in real-world scenarios, and it is natural to wonder whether pair-efficiency, the minimal efficiency requirement, is sufficient to ensure Pareto-efficiency in such cases. This paper provides a comprehensive solution to this question when agents have \textit{single-peaked} or \textit{single-dipped} preferences.
	
	Both single-peakedness and single-dippedness are natural restrictions on the preferences, where the objects are (linearly) ordered based on a certain criterion, and the preferences respect this ordering. In the case of single-peaked preferences, an agent's welfare declines as she moves away from her most preferred object (her ``peak'') in either direction according to the ordering. In the case of single-dipped preferences, an agent's welfare improves as she moves away from her least preferred object (her ``dip'') in either direction according to the ordering.
	
	Our main result (Theorem \ref{theorem equivalence}) establishes that at single-peaked or single-dipped preference profiles, pair-efficiency and Pareto-efficiency become equivalent. This means that at these structured preference profiles, any allocation that allows for an efficiency-improving trade also allows for a mutually beneficial pairwise exchange. Notably, this equivalence holds without needing to assume any other properties of the rule, highlighting the robustness and generality of the result.
	
	We further show that both the single-peaked domain and the single-dipped domain are the ``maximal'' domains where this equivalence holds (as demonstrated in Proposition \ref{proposition maximal}). However, the equivalence does not hold when agents have mixed preferences -- some with single-peaked preferences and others with single-dipped preferences (see Example \ref{example not equivalence}).

	\subsection{Related literature}
	
	\citet{ekici2024pair} is the first to explore the notion of pair-efficiency in the context of reallocating indivisible resources. He demonstrates that the \textit{top trading cycles (TTC) rule} is the only rule that satisfies strategy-proofness, individual rationality, and pair-efficiency on the unrestricted domain.\footnote{The TTC rule is first mentioned in \citet{shapley1974cores}. They attributed it to David Gale.} This strengthens the result of \citet{ma1994strategy}, which characterizes the TTC rule as the unique rule satisfying strategy-proofness, individual rationality, and Pareto-efficiency on the unrestricted domain. These two results together imply that, on the unrestricted domain, pair-efficiency is equivalent to Pareto-efficiency for any strategy-proof and individually rational rule. Later, \citet{ekici2024characterizing} provide a shorter proof of \citeauthor{ekici2024pair}'s result.
	
	Similar results are also demonstrated on the single-dipped domain. \citet{tamura2023object} characterizes the TTC rule as the only rule that satisfies strategy-proofness, individual rationality, and Pareto-efficiency on the single-dipped domain. \citet{hu2024characterization} strengthen \citeauthor{tamura2023object}'s result by showing the TTC rule as the unique rule satisfying strategy-proofness, individual rationality, and pair-efficiency on the single-dipped domain. It is worth noting that \citeauthor{hu2024characterization}'s result can be obtained as a corollary of our main result (Theorem \ref{theorem equivalence}), combined with \citeauthor{tamura2023object}'s.
	
	Unlike on the unrestricted and single-dipped domains, the TTC rule is not the only rule satisfying strategy-proofness, individual rationality, and Pareto-efficiency on the single-peaked domain. Another rule, the \textit{Crawler} \citep{bade2019matching}, also satisfies these three properties on the single-peaked domain. It is still an open problem to characterize all strategy-proof, individually rational, and Pareto-efficient rules on the single-peaked domain.

	\subsection{Organization of the paper}
	
	The paper is structured as follows: Section \ref{section preliminaries} introduces basic concepts and notations that we use throughout the paper. We describe our model, define rules, discuss their efficiency properties, and introduce the notion of single-peakedness and single-dippedness. Section \ref{section results} presents the results. Section \ref{section conclusion} concludes the paper. The appendix contains the proofs.

	\section{Preliminaries}\label{section preliminaries}

	\subsection{Model}
	
	Let $A = \{a_1, \ldots, a_n\}$ be a finite set of \textit{agents} and $H = \{h_1, \ldots, h_n\}$ be a set of equally many \textit{houses}. Each agent $a_i$ is initially endowed with house $h_i$. To avoid trivialities, we assume $n \geq 3$ throughout the paper.
	
	Agents are equipped with \textit{(strict) \textbf{preferences}} over the houses.\footnote{A \textit{(strict) preference} is a complete, asymmetric, and transitive binary relation.} For a preference $P$ and distinct $h, \tilde{h} \in H$, $h \mathrel{P} \tilde{h}$ is interpreted as ``$h$ is preferred to $\tilde{h}$ according to $P$''. For $P$, let $R$ denote the weak part of $P$, i.e., for every $h, \tilde{h} \in H$, $h \mathrel{R} \tilde{h}$ if and only if \big[$h \mathrel{P} \tilde{h}$ or $h = \tilde{h}$\big]. 
	
	Let $\mathbb{L}(H)$ denote the set of all preferences over $H$. We denote by $\mathcal{P}_a$ the \textit{set of admissible preferences} for agent $a$. Clearly, $\mathcal{P}_a \subseteq \mathbb{L}(H)$ for all $a \in A$. A \textbf{\textit{preference profile}}, denoted by $P_A = (P_{a_1}, \ldots, P_{a_n})$, is an element of the Cartesian product $\mathcal{P}_A := \mathcal{P}_{a_1} \times \cdots \times \mathcal{P}_{a_n}$, that represents a collection of preferences -- one for each agent. We call $\mathcal{P}_A$ the \textit{domain of preference profiles}, or simply the \textit{domain}. Throughout this paper, whenever we write $\mathcal{P}_A$ is the \textit{unrestricted domain}, we mean $\mathcal{P}_a = \mathbb{L}(H)$ for all $a \in A$.
	
	An \textbf{\textit{allocation}} is a one-to-one function $\mu: A \to H$.\footnote{Notice that $\mu$ is not only one-to-one, but bijective. This is because $|A| = |H|$, and any one-to-one function between two finite sets of equal size is necessarily onto.} Here, $\mu(a) = h$ means house $h$ is allocated to agent $a$ under allocation $\mu$. We denote by $\mathcal{M}$ the set of all allocations.
	
	An \textit{(allocation) \textbf{rule}} is a function $f: \mathcal{P}_A \to \mathcal{M}$. For a rule $f: \mathcal{P}_A \to \mathcal{M}$ and a preference profile $P_A \in \mathcal{P}_A$, let $f_a(P_A)$ denote the house allocated to agent $a$ by $f$ at $P_A$.

	\subsection{Restrictions on domains: Single-peakedness and single-dippedness}
	
	In many real-world situations, agents' preferences often exhibit specific restricted structures. Two such restrictions are \textit{single-peakedness} and \textit{single-dippedness}.
	
	To illustrate these concepts, consider a scenario where houses are located along a street. Suppose an agent has a particular location on the street that she favors -- perhaps due to its proximity to a park or school. The house closest to this preferred location is the agent's most desirable choice, and houses become progressively less desirable as their distance from this location increases in either direction. Such a preference structure is known as single-peaked. 
	
	Conversely, in the case of single-dipped preferences, an agent has a least preferred location -- perhaps due to the presence of a factory. The house closest to this undesirable location is the least preferred, while houses become more desirable as they are situated farther away in either direction.
	
	We now present the formal definitions of single-peakedness and single-dippedness, using the following notations.
	
	For a preference $P$, let $p(P)$ and $d(P)$ denote the most preferred (i.e., peak) and least preferred (i.e., dip) houses in $H$ according to $P$, respectively. In other words, $p(P) = h$ if and only if \big[$h \in H$ and $h \mathrel{P} \tilde{h}$ for all $\tilde{h} \in H \setminus \{h\}$\big], and $d(P) = h$ if and only if \big[$h \in H$ and $\tilde{h} \mathrel{P} h$ for all $\tilde{h} \in H \setminus \{h\}$\big].
	
	Let $\prec_H$ be a prior (linear) ordering over $H$.
	
	A preference $P$ is \textit{\textbf{single-peaked} with respect to $\prec_H$} if for every $h, \tilde{h} \in H$, $\big[p(P) \preceq_H h \prec_H \tilde{h} \mbox{ or } \tilde{h} \prec_H h \preceq_H p(P)\big]$ implies $h \mathrel{P} \tilde{h}$.\footnote{$\preceq_H$ denotes the weak part of $\prec_H$.} Let $\mathbb{SP}(\prec_H)$ denote the set of all single-peaked preferences with respect to $\prec_H$. Throughout this paper, whenever we write $\mathcal{P}_A$ is the \textit{(maximal) single-peaked domain}, we mean $\mathcal{P}_a = \mathbb{SP}(\prec_H)$ for all $a \in A$.
	
	Similarly, a preference $P$ is \textit{\textbf{single-dipped} with respect to $\prec_H$} if for every $h, \tilde{h} \in H$, $\big[d(P) \preceq_H h \prec_H \tilde{h} \mbox{ or } \tilde{h} \prec_H h \preceq_H d(P)\big]$ implies $\tilde{h} \mathrel{P} h$. Let $\mathbb{SD}(\prec_H)$ denote the set of all single-dipped preferences with respect to $\prec_H$. Throughout this paper, whenever we write $\mathcal{P}_A$ is the \textit{(maximal) single-dipped domain}, we mean $\mathcal{P}_a = \mathbb{SD}(\prec_H)$ for all $a \in A$.

	\subsection{Efficiency axioms}\label{subsection efficiency axioms}
	
	For our analysis, we consider two efficiency notions: \textit{Pareto-efficiency}, which rules out efficiency-improving trades between any group of agents, and \textit{pair-efficiency}, which only rules out efficiency-improving trades between any pair of agents. Clearly, pair-efficiency is a significant relaxation of Pareto-efficiency. The formal definitions are as follows.
	
	An allocation $\mu$ is \textit{Pareto-efficient at a preference profile $P_A \in \mathcal{P}_A$} if there exists no allocation $\nu$ such that for every $a \in A$, $\nu(a) \mathrel{R_a} \mu(a)$, and for some $b \in A$, $\nu(b) \mathrel{P_b} \mu(b)$. A rule $f: \mathcal{P}_A \to \mathcal{M}$ is \textbf{\textit{Pareto-efficient}} if for every $P_A \in \mathcal{P}_A$, $f(P_A)$ is Pareto-efficient at $P_A$.
	
	An allocation $\mu$ is \textit{pair-efficient at a preference profile $P_A \in \mathcal{P}_A$} if there do not exist distinct $a, b \in A$ such that $\mu(b) \mathrel{P_a} \mu(a)$ and $\mu(a) \mathrel{P_b} \mu(b)$. A rule $f: \mathcal{P}_A \to \mathcal{M}$ is \textbf{\textit{pair-efficient}} if for every $P_A \in \mathcal{P}_A$, $f(P_A)$ is pair-efficient at $P_A$.

	\section{Results}\label{section results}
	
	In this section, we establish the equivalence between pair-efficiency and Pareto-efficiency for preference profiles that are either single-peaked or single-dipped. Furthermore, we demonstrate that both the single-peaked domain and the single-dipped domain are the ``maximal'' domains where this equivalence holds.

	\subsection{Equivalence results}
	
	We start by presenting the main result of this paper -- the equivalence between pair-efficiency and Pareto-efficiency on the single-peaked domain, as well as on the single-dipped domain.\footnote{The concepts of single-peakedness and single-dippedness have been extended to network-based frameworks, such as trees (see, for example, \citet{demange1982single} or \citet{mandal2023compatibility} for \textit{single-peakedness on trees}, and \citet{tamura2023object} for \textit{single-dippedness on trees}). However, our result (Theorem \ref{theorem equivalence}) does not generalize to settings where preferences follow single-peakedness or single-dippedness on trees.}

	\begin{theorem}\label{theorem equivalence}
		\begin{enumerate}[(i)]
			\item\label{item equivalence single-peaked} At every single-peaked preference profile $P_A \in \mathbb{SP}^n(\prec_H)$, an allocation is Pareto-efficient if and only if it is pair-efficient.
			
			\item\label{item equivalence single-dipped} At every single-dipped preference profile $P_A \in \mathbb{SD}^n(\prec_H)$, an allocation is Pareto-efficient if and only if it is pair-efficient.
		\end{enumerate}
	\end{theorem}

	The proof of Theorem \ref{theorem equivalence} is relegated to Appendix \ref{appendix proof of theorem equivalence}; here, we provide an outline of it. The ``only-if'' part follows from the respective definitions (which holds regardless of the domain being single-peaked or single-dipped). To prove the ``if'' part, we demonstrate the contraposition. For any Pareto-inefficient allocation, an efficiency-improving trade exists. We show that in that trade, there exist two agents who prefer each other's assignments to their own, showing pair-inefficiency.
	
	What if some agents have single-peaked preferences, while others have single-dipped preferences? Does the equivalence between pair-efficiency and Pareto-efficiency still hold? Based on Theorem \ref{theorem equivalence}, one might expect that the equivalence should hold. However, that is not the case, as illustrated by the example below.

	\begin{example}\label{example not equivalence}
		Suppose $A = \{a_1, a_2, a_3\}$ and $H = \{h_1, h_2, h_3\}$ with a prior (linear) ordering $\prec_H$ such that $h_1 \prec_H h_2 \prec_H h_3$. Consider the domain $\mathcal{P}_A$ defined as follows: 
		\begin{equation*}
			\mathcal{P}_{a_1} = \mathcal{P}_{a_3} = \mathbb{SP}(\prec_H) \hspace{1 mm} \mbox{ and } \hspace{1 mm} \mathcal{P}_{a_2} = \mathbb{SD}(\prec_H).
		\end{equation*}
		We will demonstrate that pair-efficiency and Pareto-efficiency are not equivalent on $\mathcal{P}_A$. We do this by showing the existence of a preference profile at which not every pair-efficient allocation is Pareto-efficient.
		
		Consider the preference profile $P_A \in \mathcal{P}_A$ such that $P_{a_1} = h_2 h_3 h_1$, $P_{a_2} = h_3 h_1 h_2$, and $P_{a_3} = h_1 h_2 h_3$. Here, $h_1 h_2 h_3$ indicates the preference that ranks $h_1$ first, $h_2$ second, and $h_3$ third. 
		
		Now, consider the allocation $\mu = [(a_1, h_3), (a_2, h_1), (a_3, h_2)]$.\footnote{Here, $\mu(a_1) = h_3$, $\mu(a_2) = h_1$, and $\mu(a_3) = h_2$.} It is straightforward to verify that no pair of agents can benefit from trading their assignments under $\mu$ at $P_A$, and therefore, $\mu$ is pair-efficient at $P_A$. However, $\mu$ is not Pareto-efficient at $P_A$ because the allocation $\nu = [(a_1, h_2), (a_2, h_3), (a_3, h_1)]$ Pareto-dominates $\mu$ at $P_A$.\footnote{Notice that $\mu$ is individually rational at $P_A$, which implies that even under individual rationality, pair-efficiency and Pareto-efficiency are not equivalent on $\mathcal{P}_A$.} 
		
		For a more transparent illustration, the table below presents the preferences, where the pair-efficient allocation $\mu$ is highlighted in \blue{blue}, while the Pareto-dominating allocation $\nu$ is highlighted in \red{red}.
		\hfill
		$\Diamond$
		\end{example}

		\begin{table}[H]
			\centering
			\begin{tabular}{ccc}
				\hline
				$P_{a_1}$ & $P_{a_2}$ & $P_{a_3}$ \\ \hline
				\hline
				\red{$h_2$} & \red{$h_3$} & \red{$h_1$} \\
				\blue{$h_3$} & \blue{$h_1$} & \blue{$h_2$} \\
				$h_1$ & $h_2$ & $h_3$ \\
				\hline
			\end{tabular}
		\end{table}

	\subsection{Maximal domain results}
	
	In light of Theorem \ref{theorem equivalence}, one might wonder whether we can expand the domain beyond the single-peaked domain or the single-dipped domain to establish the equivalence of pair-efficiency and Pareto-efficiency, and if so, to what extent. Unfortunately, no expansion is possible. Our next result shows that both the single-peaked domain and the single-dipped domain are the ``maximal'' domains where the equivalence holds. Before presenting this result formally, it is important to recall that we only consider domains with a form of Cartesian product in this paper. Therefore, whenever $\mathcal{P}_A \supsetneq \mathbb{SP}^n(\prec_H)$, we have $\mathcal{P}_a \supseteq \mathbb{SP}(\prec_H)$ for all $a \in A$ and $\mathcal{P}_b \supsetneq \mathbb{SP}(\prec_H)$ for some $b \in A$ with $\mathcal{P}_A := \mathcal{P}_{a_1} \times \cdots \times \mathcal{P}_{a_n}$.

	\begin{proposition}\label{proposition maximal}
		\begin{enumerate}[(i)]
			\item\label{item proposition sp} Suppose $\mathcal{P}_A \supsetneq \mathbb{SP}^n(\prec_H)$. Then, there exists $P_A \in \mathcal{P}_A$ at which not every pair-efficient allocation is Pareto-efficient.
			
			\item\label{item proposition sd} Suppose $\mathcal{P}_A \supsetneq \mathbb{SD}^n(\prec_H)$. Then, there exists $P_A \in \mathcal{P}_A$ at which not every pair-efficient allocation is Pareto-efficient.
		\end{enumerate}
	\end{proposition}

	The proof of Proposition \ref{proposition maximal} is relegated to Appendix \ref{appendix proof of proposition maximal}.
	
	One important point to note is that there are sets of preference profiles that are strict supersets of the single-peaked domain (or the single-dipped domain), where the equivalence between pair-efficiency and Pareto-efficiency holds. Of course, these sets do not have the form of a Cartesian product (otherwise, we will have a contradiction to Proposition \ref{proposition maximal}). For example, consider the set $\mathbb{SP}^n(\prec_H) \cup \mathbb{SD}^n(\prec_H)$, which is a strict superset of the single-peaked domain and does not have a Cartesian product form. It follows from Theorem \ref{theorem equivalence} that at every preference profile in $\mathbb{SP}^n(\prec_H) \cup \mathbb{SD}^n(\prec_H)$, an allocation is Pareto-efficient if and only if it is pair-efficient.

	\section{Concluding remarks}\label{section conclusion}

	The findings of \citet{ma1994strategy} and \citet{ekici2024pair} suggest that pair-efficiency ensures Pareto-efficiency for any rule on the unrestricted domain, provided the rule is both strategy-proof and individually rational. In this paper, we demonstrate that pair-efficiency is equivalent to Pareto-efficiency on the single-peaked domain, as well as on the single-dipped domain, and this equivalence holds without needing to assume strategy-proofness, individual rationality, or any other properties of the rule. We further show that both the single-peaked domain and the single-dipped domain are maximal domains for this equivalence to hold.

	\appendixtocon
	\appendixtitletocon

	\renewcommand{\theequation}{\Alph{section}.\arabic{equation}}
	\renewcommand{\thetable}{\Alph{section}.\arabic{table}}
	\renewcommand{\thetheorem}{\Alph{section}.\arabic{theorem}}

	\begin{appendices}

		\section{Proof of Theorem \ref{theorem equivalence}}\label{appendix proof of theorem equivalence}
		
		\setcounter{equation}{0}
		\setcounter{table}{0}
		\setcounter{theorem}{0}

		\noindent \textbf{Proof of part \ref{item equivalence single-peaked}.} The ``only-if'' part follows from the respective definitions. We proceed to prove the ``if'' part. To do so, we show that for every $P_A \in \mathbb{SP}^n(\prec_H)$ and every $\mu \in \mathcal{M}$, $\mu$ is Pareto-efficient at $P_A$ if $\mu$ is pair-efficient at $P_A$. Assume for contradiction that this is not true; there exist a preference profile $P_A \in \mathbb{SP}^n(\prec_H)$ and an allocation $\mu \in \mathcal{M}$ such that $\mu$ is pair-efficient at $P_A$, but not Pareto-efficient. 
		
		Since $\mu$ is not Pareto-efficient at $P_A$, another allocation $\nu$ exists that Pareto-dominates $\mu$ at $P_A$. Let $\tilde{A} := \{a \in A \mid \nu(a) \mathrel{P_a} \mu(a)\}$ be the (non-empty) set of agents who prefer their assignments under $\nu$ to those under $\mu$. Since $\nu$ Pareto-dominates $\mu$ at $P_A$, by the construction of $\tilde{A}$, it follows that for every $a \in A \setminus \tilde{A}$, $\mu(a) = \nu(a)$, and consequently,
		\begin{equation}\label{equation same assignments}
			\{\mu(a) \mid a \in \tilde{A}\} = \{\nu(a) \mid a \in \tilde{A}\}.
		\end{equation}
		Label the agents in $\tilde{A}$ as $b_1, \ldots, b_m$ (i.e., $\tilde{A} = \{b_1, \ldots, b_m\}$) such that $\mu(b_1) \prec_H \cdots \prec_H \mu(b_m)$. In the rest of this proof, we will show that there exists a pair of agents $(b_l, b_{l+1})$ in $\tilde{A}$ who prefer each other's assignments under $\mu$ to their own, demonstrating pair-inefficiency of $\mu$ at $P_A$.
		
		Color each agent in $\tilde{A}$ either red or blue in the following way: color an agent $a \in \tilde{A}$ red if $\mu(a) \prec_H \nu(a)$, and color an agent $a \in \tilde{A}$ blue if $\nu(a) \prec_H \mu(a)$. Because $\mu(a) \neq \nu(a)$ for all $a \in \tilde{A}$, every agent in $\tilde{A}$ will be colored. (This colorization is for ease of presenting the logic. One can prove the theorem without relying on colorization.)

		\begin{claim}\label{claim color}
			Agent $b_1$ is red-colored, while agent $b_m$ is blue-colored.
		\end{claim}

		\begin{claimproof}[Proof of Claim \ref{claim color}.]
			Since $\mu(b_1) \neq \nu(b_1)$, the fact $\mu(b_1) \prec_H \cdots \prec_H \mu(b_m)$, together with \eqref{equation same assignments}, implies $\mu(b_1) \prec_H \nu(b_1)$, making agent $b_1$ a red-colored agent.
			
			Similarly, since $\mu(b_m) \neq \nu(b_m)$, the fact $\mu(b_1) \prec_H \cdots \prec_H \mu(b_m)$, together with \eqref{equation same assignments}, implies $\nu(b_m) \prec_H \mu(b_m)$, making agent $b_m$ a blue-colored agent. This completes the proof of Claim \ref{claim color}.
		\end{claimproof}

		Consider a pair of agents $(b_l, b_{l+1})$ in $\tilde{A}$ such that $b_l$ is a red-colored agent while $b_{l+1}$ is a blue-colored agent. Such a pair exits because of Claim \ref{claim color}.

		\begin{claim}\label{claim condition 1}
			$\mu(b_{l+1}) \mathrel{P_{b_l}} \mu(b_l)$.
		\end{claim}

		\begin{claimproof}[Proof of Claim \ref{claim condition 1}.]
			We distinguish the following two cases.
			\begin{enumerate}[(1)]
				\item\label{item claim 2 1} Suppose $p(P_{b_l}) \prec_H \mu(b_{l+1})$.
				
				Since $b_l$ is a red-colored agent, we have $\mu(b_l) \prec_H \nu(b_l)$. This, together with \eqref{equation same assignments} and the fact $\mu(b_1) \prec_H \cdots \prec_H \mu(b_m)$, implies $\mu(b_{l+1}) \preceq_H \nu(b_l)$. Because of this, and since $p(P_{b_l}) \prec_H \mu(b_{l+1})$, by single-peakedness of $P_{b_l}$, we have $\mu(b_{l+1}) \mathrel{R_{b_l}} \nu(b_l)$. This, together with the fact $\nu(b_l) \mathrel{P_{b_l}} \mu(b_l)$, implies $\mu(b_{l+1}) \mathrel{P_{b_l}} \mu(b_l)$.

				\item\label{item claim 2 2} Suppose $\mu(b_{l+1}) \preceq_H p(P_{b_l})$.
				
				Since $\mu(b_{l+1}) \preceq_H p(P_{b_l})$, the fact $\mu(b_1) \prec_H \cdots \prec_H \mu(b_m)$ implies $\mu(b_l) \prec_H \mu(b_{l+1}) \preceq_H p(P_{b_l})$. Because of this, and since $P_{b_l}$ is single-peaked, we have $\mu(b_{l+1}) \mathrel{P_{b_l}} \mu(b_l)$.
			\end{enumerate}
			
			Since Cases \ref{item claim 2 1} and \ref{item claim 2 2} are exhaustive, this completes the proof of Claim \ref{claim condition 1}.
		\end{claimproof}

		\begin{claim}\label{claim condition 2}
			$\mu(b_l) \mathrel{P_{b_{l+1}}} \mu(b_{l+1})$.
		\end{claim}

		\begin{claimproof}[Proof of Claim \ref{claim condition 2}.]
			We distinguish the following two cases.
			\begin{enumerate}[(1)]
				\item\label{item claim 3 1} Suppose $p(P_{b_{l+1}}) \preceq_H \mu(b_l)$.
				
				Since $p(P_{b_{l+1}}) \preceq_H \mu(b_l)$, the fact $\mu(b_1) \prec_H \cdots \prec_H \mu(b_m)$ implies $p(P_{b_{l+1}}) \preceq_H \mu(b_l) \prec_H \mu(b_{l+1})$. Because of this, and since $P_{b_{l+1}}$ is single-peaked, we have $\mu(b_l) \mathrel{P_{b_{l+1}}} \mu(b_{l+1})$.

				\item\label{item claim 3 2} Suppose $\mu(b_l) \prec_H p(P_{b_{l+1}})$.
				
				Since $b_{l+1}$ is a blue-colored agent, we have $\nu(b_{l+1}) \prec_H \mu(b_{l+1})$. This, together with \eqref{equation same assignments} and the fact $\mu(b_1) \prec_H \cdots \prec_H \mu(b_m)$, implies $\nu(b_{l+1}) \preceq_H \mu(b_l)$. Because of this, and since $\mu(b_l) \prec_H p(P_{b_{l+1}})$, by single-peakedness of $P_{b_l}$, we have $\mu(b_l) \mathrel{R_{b_{l+1}}} \nu(b_{l+1})$. This, together with the fact $\nu(b_{l+1}) \mathrel{P_{b_{l+1}}} \mu(b_{l+1})$, implies $\mu(b_l) \mathrel{P_{b_{l+1}}} \mu(b_{l+1})$.
			\end{enumerate}
			
			Since Cases \ref{item claim 3 1} and \ref{item claim 3 2} are exhaustive, this completes the proof of Claim \ref{claim condition 2}.
		\end{claimproof}

		However, Claims \ref{claim condition 1} and \ref{claim condition 2} together contradict that $\mu$ is pair-efficient at $P_A$, which completes the proof of part \ref{item equivalence single-peaked} of Theorem \ref{theorem equivalence}.\bigskip

		\noindent \textbf{Proof of part \ref{item equivalence single-dipped}.} The ``only-if'' part follows from the respective definitions. We proceed to prove the ``if'' part. To do so, we show that for every $P_A \in \mathbb{SD}^n(\prec_H)$ and every $\mu \in \mathcal{M}$, $\mu$ is Pareto-efficient at $P_A$ if $\mu$ is pair-efficient at $P_A$. Assume for contradiction that this is not true; there exist a preference profile $P_A \in \mathbb{SD}^n(\prec_H)$ and an allocation $\mu \in \mathcal{M}$ such that $\mu$ is pair-efficient at $P_A$, but not Pareto-efficient. 
		
		Since $\mu$ is not Pareto-efficient at $P_A$, another allocation $\nu$ exists that Pareto-dominates $\mu$ at $P_A$. Let $\tilde{A} := \{a \in A \mid \nu(a) \mathrel{P_a} \mu(a)\}$ be the (non-empty) set of agents who prefer their assignments under $\nu$ to those under $\mu$. Since $\nu$ Pareto-dominates $\mu$ at $P_A$, by the construction of $\tilde{A}$, it follows that for every $a \in A \setminus \tilde{A}$, $\mu(a) = \nu(a)$, and consequently,
		\begin{equation}\label{equation same assignments sd}
			\{\mu(a) \mid a \in \tilde{A}\} = \{\nu(a) \mid a \in \tilde{A}\}.
		\end{equation}
		Label the agents in $\tilde{A}$ as $b_1, \ldots, b_m$ (i.e., $\tilde{A} = \{b_1, \ldots, b_m\}$) such that $\mu(b_1) \prec_H \cdots \prec_H \mu(b_m)$. In the rest of this proof, we will show that the agents $b_1$ and $b_m$ prefer each other's assignments under $\mu$ to their own, demonstrating pair-inefficiency of $\mu$ at $P_A$.

		\begin{claim}\label{claim condition sd 1}
			$\mu(b_m) \mathrel{P_{b_1}} \mu(b_1)$.
		\end{claim}

		\begin{claimproof}[Proof of Claim \ref{claim condition sd 1}.]
			Since $\mu(b_1) \neq \nu(b_1)$, the fact $\mu(b_1) \prec_H \cdots \prec_H \mu(b_m)$, together with \eqref{equation same assignments sd}, implies $\mu(b_1) \prec_H \nu(b_1) \preceq_H \mu(b_m)$. Furthermore, since $\nu(b_1) \mathrel{P_{b_1}} \mu(b_1)$ and $\mu(b_1) \prec_H \nu(b_1)$, by single-dippedness of $P_{b_1}$, we have $d(P_{b_1}) \prec_H \nu(b_1)$. Because of this, and since $\nu(b_1) \preceq_H \mu(b_m)$, by single-dippedness of $P_{b_1}$, we have $\mu(b_m) \mathrel{R_{b_1}} \nu(b_1)$. This, together with the fact $\nu(b_1) \mathrel{P_{b_1}} \mu(b_1)$, implies $\mu(b_m) \mathrel{P_{b_1}} \mu(b_1)$. This completes the proof of Claim \ref{claim condition sd 1}.
		\end{claimproof}

		\begin{claim}\label{claim condition sd 2}
			$\mu(b_1) \mathrel{P_{b_m}} \mu(b_m)$.
		\end{claim}

		\begin{claimproof}[Proof of Claim \ref{claim condition sd 2}.]
			Since $\mu(b_m) \neq \nu(b_m)$, the fact $\mu(b_1) \prec_H \cdots \prec_H \mu(b_m)$, together with \eqref{equation same assignments sd}, implies $\mu(b_1) \preceq_H \nu(b_m) \prec_H \mu(b_m)$. Furthermore, since $\nu(b_m) \mathrel{P_{b_m}} \mu(b_m)$ and $\nu(b_m) \prec_H \mu(b_m)$, by single-dippedness of $P_{b_m}$, we have $\nu(b_m) \prec_H d(P_{b_m})$. Because of this, and since $\mu(b_1) \preceq_H \nu(b_m)$, by single-dippedness of $P_{b_m}$, we have $\mu(b_1) \mathrel{R_{b_m}} \nu(b_m)$. This, together with the fact $\nu(b_m) \mathrel{P_{b_m}} \mu(b_m)$, implies $\mu(b_1) \mathrel{P_{b_m}} \mu(b_m)$. This completes the proof of Claim \ref{claim condition sd 2}.
		\end{claimproof}

		However, Claims \ref{claim condition sd 1} and \ref{claim condition sd 2} together contradict that $\mu$ is pair-efficient at $P_A$, which completes the proof of part \ref{item equivalence single-dipped} of Theorem \ref{theorem equivalence}.
		\hfill
		\qed

		\section{Proof of Proposition \ref{proposition maximal}}\label{appendix proof of proposition maximal}
		
		\setcounter{equation}{0}
		\setcounter{table}{0}
		\setcounter{theorem}{0}

		\noindent \textbf{Proof of part \ref{item proposition sp}.} Since $\mathcal{P}_A \supsetneq \mathbb{SP}^n(\prec_H)$, there exists an agent $a \in A$ such that $\mathcal{P}_a \supsetneq \mathbb{SP}(\prec_H)$. Consider a preference $P_a$ of agent $a$ such that $P_a \in \mathcal{P}_a \setminus \mathbb{SP}(\prec_H)$. Let $p(P_a) = h$. (Note that house $h$ does not necessarily have to be the initial endowment of agent $a$.) Since $P_a \in \mathcal{P}_a \setminus \mathbb{SP}(\prec_H)$, there exist distinct $h', \tilde{h} \in H \setminus \{h\}$ with $h \mathrel{P_a} \tilde{h} \mathrel{P_a} h'$ where
		\begin{equation*}
			h \prec_H h' \prec_H \tilde{h} \hspace{2 mm} \mbox{ or } \hspace{2 mm} \tilde{h} \prec_H h' \prec_H h.
		\end{equation*}
		
		We now construct a preference profile at which not every pair-efficient allocation is Pareto-efficient. Fix two distinct agents $a', \tilde{a} \in A \setminus \{a\}$ and a bijection $\beta : A \setminus \{a, a', \tilde{a}\} \to H \setminus \{h, h', \tilde{h}\}$. (Note that house $h'$ does not need to be the initial endowment of agent $a'$, nor does house $\tilde{h}$ have to be the initial endowment of agent $\tilde{a}$.) Consider the preference profile $P^*_A \in \mathcal{P}_A$ such that
		\begin{enumerate}[(a)]
			\item $P^*_a = P_a$,
			
			\item $P^*_{a'} \in \mathbb{SP}(\prec_H)$ with $h' \mathrel{P^*_{a'}} h \mathrel{P^*_{a'}} \tilde{h}$,
			
			\item $P^*_{\tilde{a}} \in \mathbb{SP}(\prec_H)$ with $\tilde{h} \mathrel{P^*_{\tilde{a}}} h' \mathrel{P^*_{\tilde{a}}} h$, and 
			
			\item for every $b \in A \setminus \{a, a', \tilde{a}\}$, $P^*_b \in \mathbb{SP}(\prec_H)$ with $p(P^*_b) = \beta(b)$.
		\end{enumerate}
		
		Consider the allocation $\mu = [(a, \tilde{h}), (a', h), (\tilde{a}, h'), (b, \beta(b)) \mbox{ for all } b \in A \setminus \{a, a', \tilde{a}\}]$. We complete the proof by demonstrating that $\mu$ is pair-efficient at $P^*_A$, but not Pareto-efficient. To provide clear intuition for this claim, we first present a table summarizing the preferences. In the table, the pair-efficient allocation $\mu$ is highlighted in \blue{blue}, while the efficiency-improving trade is highlighted in \red{red}.
		\begin{table}[H]
			\centering
			\begin{tabular}{ccccc}
				\hline
				$P^*_{a}$ & $P^*_{a'}$ & $P^*_{\tilde{a}}$ & $\ldots$ & $P^*_b$ \\ \hline
				\hline
				\red{$h$} & $\vdots$ & $\vdots$ & $\ldots$ & \blue{$\beta(b)$} \\
				$\vdots$ & \red{$h'$} & \red{$\tilde{h}$} & $ $ & $\vdots$ \\
				\blue{$\tilde{h}$} & $\vdots$ & $\vdots$ & $ $ & $ $ \\
				$\vdots$ & \blue{$h$} & \blue{$h'$} & $ $ & $ $ \\
				$h'$ & $\vdots$ & $\vdots$ & $ $ & $ $ \\
				$\vdots$ & $\tilde{h}$ & $h$ & $ $ & $ $ \\
				$ $ & $\vdots$ & $\vdots$ & $ $ & $ $ \\
				\hline
			\end{tabular}
		\end{table}
		
		Now, we formally explain why $\mu$ is pair-efficient at $P^*_A$, but not Pareto-efficient. Since every agent in $A \setminus \{a, a', \tilde{a}\}$ is assigned their most preferred house under $\mu$ at $P^*_A$, no agent in $A \setminus \{a, a', \tilde{a}\}$ will be part of a mutually beneficial pairwise exchange even if such a pairwise exchange exists. Furthermore, it is straightforward to verify that no pair of agents in $\{a, a', \tilde{a}\}$ can benefit from trading their assignments under $\mu$ at $P^*_A$. Combining these two observations, it follows that $\mu$ is pair-efficient at $P^*_A$. However, $\mu$ is not Pareto-efficient at $P^*_A$ because the allocation $\nu = [(a, h), (a', h'), (\tilde{a}, \tilde{h}), (b, \beta(b)) \mbox{ for all } b \in A \setminus \{a, a', \tilde{a}\}]$ Pareto-dominates $\mu$ at $P^*_A$. This completes the proof of part \ref{item proposition sp} of Proposition \ref{proposition maximal}.\bigskip

		\noindent \textbf{Proof of part \ref{item proposition sd}.} Since $\mathcal{P}_A \supsetneq \mathbb{SD}^n(\prec_H)$, there exists an agent $a \in A$ such that $\mathcal{P}_a \supsetneq \mathbb{SD}(\prec_H)$. Consider a preference $P_a$ of agent $a$ such that $P_a \in \mathcal{P}_a \setminus \mathbb{SD}(\prec_H)$. Let $d(P_a) = h$. (Note that house $h$ does not necessarily have to be the initial endowment of agent $a$.) Since $P_a \in \mathcal{P}_a \setminus \mathbb{SD}(\prec_H)$, there exist distinct $h', \tilde{h} \in H \setminus \{h\}$ with $h' \mathrel{P_a} \tilde{h} \mathrel{P_a} h$ where
		\begin{equation*}
			h \prec_H h' \prec_H \tilde{h} \hspace{2 mm} \mbox{ or } \hspace{2 mm} \tilde{h} \prec_H h' \prec_H h.
		\end{equation*}
		In the rest of this proof, we will construct a preference profile at which not every pair-efficient allocation is Pareto-efficient. We distinguish the following two cases.\medskip

		\noindent\textbf{\textsc{Case} 1}: Suppose $h \prec_H h' \prec_H \tilde{h}$.
		
		First, we construct the required preference profile. Fix two distinct agents $a', \tilde{a} \in A \setminus \{a\}$ and a bijection $\beta : A \setminus \{a, a', \tilde{a}\} \to H \setminus \{h, h', \tilde{h}\}$. (Note that house $h'$ does not need to be the initial endowment of agent $a'$, nor does house $\tilde{h}$ have to be the initial endowment of agent $\tilde{a}$.) Consider three preferences $P^1, P^2, P^3 \in \mathbb{SD}(\prec_H)$ such that
		\begin{enumerate}[(a)]
			\item $h_i \mathrel{P^1} h_j$ for all $h_i, h_j \in H$ with $h_i \prec_H h_j$,
			
			\item $h_j \mathrel{P^2} h_i$ for all $h_i, h_j \in H$ with $h_i \prec_H h_j$, and
			
			\item $d(P^3) = h'$, and $h_k \mathrel{P^3} h_i \mathrel{P^3} h_j$ for all $h_i, h_j, h_k \in H$ with $h_i \preceq_H h \prec_H h_j \prec_H \tilde{h} \preceq_H h_k$.
		\end{enumerate}
		Consider the preference profile $P^*_A \in \mathcal{P}_A$ such that
		\begin{equation*}
			\begin{aligned}
				& P^*_a = P_a, \hspace{2 mm} P^*_{a'} = P^3, \hspace{2 mm} P^*_{\tilde{a}} = P^1,\\
				& P^*_b = P^1 \mbox{ for all } b \in A \setminus \{a, a', \tilde{a}\} \mbox{ with } \beta(b) \prec_H \tilde{h}, \mbox{ and }\\
				& P^*_b = P^2 \mbox{ for all } b \in A \setminus \{a, a', \tilde{a}\} \mbox{ with } \tilde{h} \prec_H \beta(b).
			\end{aligned}
		\end{equation*}
		
		Consider the allocation $\mu = [(a, \tilde{h}), (a', h), (\tilde{a}, h'), (b, \beta(b)) \mbox{ for all } b \in A \setminus \{a, a', \tilde{a}\}]$. We complete the proof by demonstrating that $\mu$ is pair-efficient at $P^*_A$, but not Pareto-efficient. It follows from the construction of $P^*_A$ and $\mu$ that no agent in $A \setminus \{a, a', \tilde{a}\}$ will be part of a mutually beneficial pairwise exchange even if such a pairwise exchange exists. Furthermore, it is straightforward to verify that no pair of agents in $\{a, a', \tilde{a}\}$ can benefit from trading their assignments under $\mu$ at $P^*_A$. Combining these two observations, it follows that $\mu$ is pair-efficient at $P^*_A$. However, $\mu$ is not Pareto-efficient at $P^*_A$ because the allocation $\nu = [(a, h'), (a', \tilde{h}), (\tilde{a}, h), (b, \beta(b)) \mbox{ for all } b \in A \setminus \{a, a', \tilde{a}\}]$ Pareto-dominates $\mu$ at $P^*_A$.\medskip

		\noindent\textbf{\textsc{Case} 2}: Suppose $\tilde{h} \prec_H h' \prec_H h$.
		
		First, we construct the required preference profile. Fix two distinct agents $a', \tilde{a} \in A \setminus \{a\}$ and a bijection $\beta : A \setminus \{a, a', \tilde{a}\} \to H \setminus \{h, h', \tilde{h}\}$. (Note that house $h'$ does not need to be the initial endowment of agent $a'$, nor does house $\tilde{h}$ have to be the initial endowment of agent $\tilde{a}$.) Consider three preferences $P^1, P^2, P^3 \in \mathbb{SD}(\prec_H)$ such that
		\begin{enumerate}[(a)]
			\item $h_i \mathrel{P^1} h_j$ for all $h_i, h_j \in H$ with $h_i \prec_H h_j$,
			
			\item $h_j \mathrel{P^2} h_i$ for all $h_i, h_j \in H$ with $h_i \prec_H h_j$, and
			
			\item $d(P^3) = h'$, and $h_i \mathrel{P^3} h_k \mathrel{P^3} h_j$ for all $h_i, h_j, h_k \in H$ with $h_i \preceq_H \tilde{h} \prec_H h_j \prec_H h \preceq_H h_k$.
		\end{enumerate}
		Consider the preference profile $P^*_A \in \mathcal{P}_A$ such that
		\begin{equation*}
			\begin{aligned}
				& P^*_a = P_a, \hspace{2 mm} P^*_{a'} = P^3, \hspace{2 mm} P^*_{\tilde{a}} = P^2,\\
				& P^*_b = P^1 \mbox{ for all } b \in A \setminus \{a, a', \tilde{a}\} \mbox{ with } \beta(b) \prec_H \tilde{h}, \mbox{ and }\\
				& P^*_b = P^2 \mbox{ for all } b \in A \setminus \{a, a', \tilde{a}\} \mbox{ with } \tilde{h} \prec_H \beta(b).
			\end{aligned}
		\end{equation*}
		
		Consider the allocation $\mu = [(a, \tilde{h}), (a', h), (\tilde{a}, h'), (b, \beta(b)) \mbox{ for all } b \in A \setminus \{a, a', \tilde{a}\}]$. We complete the proof by demonstrating that $\mu$ is pair-efficient at $P^*_A$, but not Pareto-efficient. It follows from the construction of $P^*_A$ and $\mu$ that no agent in $A \setminus \{a, a', \tilde{a}\}$ will be part of a mutually beneficial pairwise exchange even if such a pairwise exchange exists. Furthermore, it is straightforward to verify that no pair of agents in $\{a, a', \tilde{a}\}$ can benefit from trading their assignments under $\mu$ at $P^*_A$. Combining these two observations, it follows that $\mu$ is pair-efficient at $P^*_A$. However, $\mu$ is not Pareto-efficient at $P^*_A$ because the allocation $\nu = [(a, h'), (a', \tilde{h}), (\tilde{a}, h), (b, \beta(b)) \mbox{ for all } b \in A \setminus \{a, a', \tilde{a}\}]$ Pareto-dominates $\mu$ at $P^*_A$.\medskip

		Since Cases 1 and 2 are exhaustive, this completes the proof of part \ref{item proposition sd} of Proposition \ref{proposition maximal}.
		\hfill
		\qed

	\end{appendices}

	\section*{Declarations}

	\paragraph{Declaration of competing interest}
	
	The author has no competing interests to declare that are relevant to the content of this article.

	\paragraph{Data availability}
	
	No data was used for the research described in the article.

	\paragraph{Funding}
	
	This research did not receive any specific grant from funding agencies in the public, commercial, or not-for-profit sectors.

	\setcitestyle{numbers}
	\bibliographystyle{plainnat}
	\bibliography{mybib}

\end{document}